\documentclass[final,5p,times,twocolumn]{elsarticle}

\usepackage{amssymb}
\usepackage{amsmath}
\usepackage{physics}
\usepackage{dcolumn}
\usepackage{booktabs}
\usepackage{color}
\usepackage{hyperref}
\usepackage{setspace}
\usepackage{subfigure}

\begin{document}

\begin{frontmatter}



\title{Light Rings, Accretion Disks and Shadows of Hayward Boson Stars}

\author[a]{Zhen-Hua Zhao \corref{cor1}} 
\ead{zhaozhh78@sdust.edu.cn}
\author[a]{Yi-Ning Gu}
\author[b]{Shu-Cong Liu}
\author[a,b]{Zi-Qian Liu}
\author[b]{Yong-Qiang Wang \corref{cor2}}
\ead{yqwang@lzu.edu.cn}
\cortext[cor1]{corresponding author}
\cortext[cor2]{corresponding author}

 \affiliation[a]{organization={Department of Applied Physics, Shandong University of Science and Technology},
             city={Qingdao},
             postcode={266590},
             country={China}}
\affiliation[b]{organization={Lanzhou Center for Theoretical Physics,
School of Physical Science and Technology, Lanzhou University},
             city={Lanzhou},
             postcode={730000},
             country={China}}
%
%

\begin{abstract}

In this paper,  we investigate the Einstein-Hayward { gravity  coupled} to a complex scalar field without self-interaction. Using numerical methods, we construct a class of  Hayward boson star solutions and examine their fundamental properties as well as the optical appearance of the {  accretion  disk. Our results show  that in the frozen state,} both the { quasi-horizon} radius and the light ring radii increase with the magnetic monopole charge. Furthermore,  using ray-tracing method, we find that for { non-frozen} states, the absence of an { quasi-horizon} results in the appearance of multiple photon rings within the  shadow region of the accretion disks. In contrast, for frozen states, the presence of a { quasi-horizon} causes their images to resemble those of Schwarzschild black holes, with no additional photon rings appearing.  
\end{abstract}




\end{frontmatter}



\section{Introduction}\label{sec:intro}

The general theory of relativity, proposed by Einstein in 1916 \cite{Einstein1916}, has been remarkably successful in describing gravitational phenomena, especially in extreme conditions such as strong gravitational fields. Notable examples include the  Schwarzschild  black hole \cite{Schwarzschild1916}, Kerr black hole \cite{Kerr1963} and Kerr-Newman black hole \cite{NewmanCouchChinnapared1965}. However, the issue of singularities remains unresolved. Singularities are not a reflection of the real physical world but rather a limitation of existing theories when describing the behavior of matter and spacetime under extreme conditions.

To address the singularity problem, regular black holes are proposed as theoretical models to avoid singularity formation. Early attempts to regularize black hole metrics were made by Shirokov (1948) \cite{Shirokov1948}, Duan (1954) \cite{Duan1954}, Sakharov \cite{Sakharov1966} (1966), and Gliner (1966) \cite{Gliner1966}. In 1968, Bardeen proposed a widely accepted model of regular black holes \cite{Bardeen1968}. Further in 2006, Hayward introduced another model of regular black holes \cite{Hayward2006}.

However, obtaining exact solutions for regular black holes by solving the Einstein equations remains highly challenging.  It was not until 1999 that Ayón-Beato and García \cite{Ayon-BeatoGarcia1998} proposed the first exact regular black hole solution using nonlinear electrodynamics coupled to general relativity. Subsequently, they  derived the exact solution for the Bardeen black hole \cite{Ayon-BeatoGarcia2000}, interpreting it as a  magnetic monopole. Further, the exact solution Hayward black hole is obtained by Fan and Wang in 2016 \cite{FanWang2016}.

{
In addition to the regular black hole, boson stars \cite{Wheeler1955,Kaup1968,RuffiniBonazzola1969}, as a class of compact objects without  central singularities and event horizons, can resemble black holes in many observational features, making them an important subject of theoretical research. In recent years, theoretical studies on their observational characteristics have advanced significantly. 
For example, Vincent et al. simulated the image of a possible boson star at the Galactic center\cite{VincentMelianiGrandclement2016}; Cunha et al. calculated the gravitational lensing effects and dynamical features of ultra-compact boson stars \cite{CunhaFontHerdeiro2017}; and Olivares et al. systematically explored how to observationally distinguish accreting boson stars from black holes \cite{OlivaresYounsiFromm2020}. Subsequent research has further focused on specific observational features, including hot spots   near boson stars \cite{RosaGarciaVincent2022,RosaMacedoRubiera-Garcia2023}, shadows \cite{RosaRubiera-Garcia2022} and the relativistic broadening of emission lines  of accretion disks \cite{RosaPellePerez2024}. Moreover, related studies have expanded into different theoretical frameworks and model comparisons, including static solutions and their images in Einstein-Friedberg-Lee-Sirlin theory \cite{deSaLimaHerdeiro2024}, comparisons of magnetized thick disks around black holes and boson stars \cite{GjorgjieskiKunzNedkova2024}, and optical images of mini-boson stars in Palatini f(R) gravity theory \cite{ZengYangHuang2025}. Recent work has also meticulously examined the observational features of massive boson stars under thin-disk accretion \cite{LiWuHe2025}, images of solitonic boson stars illuminated by different accreting materials \cite{HeLiYang2025}, and the lensing effects and ring structures of boson stars with parity-odd rotation \cite{HuangLiuZhang2025}. These studies systematically characterize the possible observational signatures of boson stars from multiple dimensions, providing  theoretical foundations for testing or constraining such horizonless compact objects using high-resolution observational data.

In this paper, we focus on a boson star with magnetic monopole charge based on the Hayward spacetime \cite{Hayward2006,FanWang2016} (referred to as the Hayward boson star \cite{YueWang2025}). This boson star has a distinctive feature: when the magnetic monopole charge exceeds a certain critical value, the oscillation frequency of the complex scalar field in the model can approach infinitely close to zero ( not strictly zero \cite{FriedbergLeePang1987,Hod2018}). Based on the work of Yue and Wang \cite{YueWang2025}, we further explored the influence of magnetic monopole charges on the model and compared the differences in images of the accretion disk under frozen and non-frozen states of Hayward boson star by using the ray-tracing method.

}

The organization of this paper is as follows. In Sec. \ref{model}, the model of  { Hayward boson stars with a magnetic monopole  } is presented. In Sec. \ref{sect3}, the numerical scheme is outlined. In Sec. \ref{sect4}, the numerical results and discussion are presented in {three} topics. In Sec. \ref{Conc}, some conclusions are presented.

\section{Introduction to the Model}\label{model}

The model we study is constructed within the theoretical framework coupling gravity with nonlinear electromagnetic fields. Research on nonlinear electromagnetic theory can be traced back to the pioneering work of Born and Infeld (BI) in field theory \cite{BornInfeld1934}, as well as the foundational studies of Heisenberg and Euler (HE) in quantum electrodynamics \cite{HeisenbergEuler1936}. Since then, the combination of nonlinear electromagnetic fields with Einsteinian gravity has been widely applied in black hole physics research, exhibiting methodological diversity: one line of research follows the BI tradition, adopting closed-form nonlinear electromagnetic field Lagrangians \cite{Ayon-BeatoGarcia1998,Kruglov2017,RahmatovZahidKhan2025}; another follows the HE approach, expressing the Lagrangian as polynomial forms of electromagnetic field strength \cite{Soleng1995,ZhangYangZou2015,LiuMaiLi2019}; some works consider both scenarios \cite{deOliveira1994,YuGao2020}; certain studies do not require explicit specification of the Lagrangian form \cite{Bronnikov2001,BalartVagenas2014,ChenDeFeliceTsujikawa2025}; additionally, there are investigations exploring Yang-Mills field cases \cite{DuGu2024}.

The model we investigate  is  based on the Hayward black hole solution \cite{FanWang2016, Fan2017} with nonlinear electromagnetic field, with the inclusion of a complex scalar field to explore boson star solutions. When this complex scalar field is neglected, the model reduces to the original Hayward black hole solution. Therefore, we refer to the studied object as a Hayward boson star. Here, we work in   natural units ($\hbar = c = 1$), and the action is given by:
\begin{equation}\label{action}  
S = \int \sqrt{-g} \, d^4x \left( \frac{R}{4\kappa} + \mathcal{L}^{(1)} + \mathcal{L}^{(2)} \right),  
\end{equation}
with $\kappa = 4\pi G$, $G$ is the gravitational constant.  
The Lagrangian densities are defined as:
\begin{eqnarray}  
\mathcal{L}^{(1)} &=& -\frac{3}{2s} \frac{(2\hat{q}^2  \hat{\mathcal{F})}^{3/2}}{\left(1 + (2\hat{q}^2  \hat{\mathcal{F}})^{3/4}\right)^2}, \label{vectA}\\  
\mathcal{L}^{(2)} &=& -\nabla_\mu \Phi^* \nabla^\mu \Phi - U(\Phi, \Phi^*), 
\end{eqnarray}
where,  
 $\hat{\mathcal{F}} = \frac{ 1}{4} \hat F_{\mu\nu} \hat F^{\mu\nu}$, $\hat F_{\mu\nu} = \partial_\mu \hat A_\nu - \partial_\nu \hat A_\mu$,   $\hat{q} = q\sqrt{G}$,  $q$ denotes the  magnetic monopole charge,  and parameter $s$ is free. It is worth noting that the electromagnetic action \eqref{vectA} used in this paper is similar to that in \cite{Ayon-BeatoGarcia2000}, but differs from the one adopted in \cite{Fan2017}.
 
The  {{potential function}} of  the complex scalar field is
\[  
U(\Phi\Phi^*) = \mu^2 \Phi \Phi^*,  
\]
where $\Phi$ is a complex scalar field with conjugate $\Phi^*$ and $\mu$ is the mass parameter.

Varying the action (\ref{action}) with respect to $g_{\mu\nu}$ and $\Phi$ yields the equations of motion:
\begin{eqnarray}\label{eq:EKG1}  
R_{\mu\nu} - \frac{1}{2} g_{\mu\nu} R - 2\kappa \left(T^{(1)}_{\mu\nu} + T^{(2)}_{\mu\nu}\right) &=& 0, \\  
\nabla_\mu \left( \frac{\partial \mathcal{L}^{(1)}}{\partial {\hat{\mathcal{F}} }} {\hat{F}^{\mu\nu} }\right) &=& 0, \\  
\Box \Phi - \dot{U}(\Phi\Phi^*) \Phi &=& 0.  
\end{eqnarray}
Here, $\dot{U}(\Phi\Phi^*) = \mu^2$  and the energy-momentum tensors are:
\begin{equation}  
T^{(1)}_{\mu\nu} = -\frac{\partial \mathcal{L}^{(1)}}{\partial {\hat{\mathcal{F}}}} { \hat{F}_{\mu\rho}  \hat{F}_\nu^{\;\rho} }+ g_{\mu\nu} \mathcal{L}^{(1)},  
\end{equation}
\begin{eqnarray}  
 &  & T^{(2)}_{\mu\nu} =\partial_\mu \Phi^* \partial_\nu \Phi + \partial_\nu \Phi^* \partial_\mu \Phi \nonumber\\
&  &  - g_{\mu\nu} \left[ \frac{1}{2} g^{\lambda\rho} \left( \partial_\lambda \Phi^* \partial_\rho \Phi + \partial_\rho \Phi^* \partial_\lambda \Phi \right) + U(\Phi\Phi^*) \right].  
\end{eqnarray}

We consider a static, spherically symmetric spacetime with the metric:
\begin{equation}\label{equ10}  
ds^2 = -N(r)\sigma^2(r) dt^2 + \frac{dr^2}{N(r)} + r^2 \left( d\theta^2 + \sin^2\theta \, d\varphi^2 \right),  
\end{equation}
where 
\[  
{ N(r) = 1 - \frac{2G m(r)}{r} }.  
\]
{{Additionally,   the  ansatzes for the electromagnetic field and the scalar field are considered,
\begin{equation}\label{equ11}  
 \hat A_\mu = \{0, 0, 0, \hat {q} \cos (\theta) \}, \ \Phi = \frac{\phi(r)}{\sqrt{2}} e^{-i\omega t},  
\end{equation}
with $\phi(r)$ being a real scalar field and $\omega$ the oscillation frequency.}}

Substituting (\ref{equ10}) and (\ref{equ11}) into (\ref{eq:EKG1}) yields the following ordinary differential equations:

\begin{align}
\phi'' + \left( \frac{2}{r} + \frac{N'}{N} + \frac{\sigma'}{\sigma} \right) \phi' + \left( \frac{\omega^2}{N \sigma^2} - \frac{\mu^2}{2} \right) \frac{\phi}{N} & \!\!=\! 0, \label{ode11} \\  
N' + \frac{\kappa \omega^2 r \phi^2}{N \sigma^2} + \kappa r N \phi'^2 + \frac{N - 1}{r} \nonumber \\
{ \qquad +\frac{3 \kappa \hat{q}^6 r}{s (\hat{q}^3 + r^3)^2} +   \kappa \mu^2 \phi^2 r } &  \!\!=\! 0,  \label{ode12} \\  
\frac{\sigma'}{\sigma} - \kappa r \left( \phi'^2 + \frac{\omega^2 \phi^2}{N^2 \sigma^2} \right) & \!\!=\! 0, \label{ode13}  
\end{align}
where 
$${ N' = -\frac{2G m'(r)}{r} + \frac{2G m(r)}{r^2} , }$$ 
and ``$'$'' is used to  denote differentiation with respect to $r$.

To ensure regularity and asymptotic flatness, the following boundary conditions are proposed,
\begin{equation}\label{equ19}  
m(0) = 0, \quad \sigma(\infty) = 1,   
\end{equation}
\begin{equation}\label{equ20}  
\phi(\infty) = 0, \quad \left. \frac{d\phi}{dr} \right|_{r=0} = 0.  
\end{equation}

Moreover, the Arnowitt-Deser-Misner (ADM) mass is given by the asymptotic limit of the solution $ m(r) $ as $ r \to \infty $, with $ m(\infty) \equiv M $.

{ It should be noted  although the complex scalar field itself oscillates in time, the energy-momentum tensors $T^{(1)}_{\mu\nu}$, $T^{(2)}_{\mu\nu}$ and all physical observables derived from it are {time-independent}. This occurs because the stress-energy tensor depends only on the modulus of the scalar field $\phi$, which is time-independent in our ansatz. Therefore, the metric and gauge field remain strictly static, and there is no scalar radiation associated with this solution. }

\section{Numerical  scheme} \label{sect3}

To facilitate our numerical analysis, we introduce dimensionless variables through the following scaling transformations:

\begin{equation}\label{scaling}
\begin{aligned}
r \to \frac{\tilde{r}}{\mu}, \  \hat{q} \to \frac{\tilde{q}}{\mu}, \  \omega \to \tilde{\omega}\mu,  \\
s \to \frac{\kappa \tilde{s}}{\mu^2}, \ m \to \frac{\tilde{m}}{G\mu}, \ \phi \to \frac{\tilde{\phi}}{\sqrt{\kappa}}, \  M \to \frac{\tilde{M}}{G\mu}. 
\end{aligned}
\end{equation}
After applying these  transformations, the form of  field equations \eqref{ode11}--\eqref{ode13} can be simplified  by setting:
\begin{equation}
\mu \to 1, \quad G \to 1, \quad \text{and} \quad \kappa \to 1.
\end{equation}

In addition, for the convenience of numerical calculations, a compactified radial coordinate is introduced:
\begin{equation}\label{x_r}
x = \frac{\tilde{r}}{1 + \tilde{r}},
\end{equation}
which maps the physical domain $r \in [0, \infty)$ to the computational domain $x \in [0,1]$. This transformation yields the following correspondence: $x=0$ represents the origin (central point of the solution), while $x=1$ corresponds to spatial infinity.

Finally, the boundary value problem defined by equations \eqref{ode11}--\eqref{ode13} and the boundary conditions \eqref{equ19} and \eqref{equ20} is solved using the open-source finite element computing software FEniCSx (version 0.8) \cite{BarattaDeanDokken2023,ScroggsDokkenRichardson2022,ScroggsBarattaRichardson2022,AlnaesLoggOlgaard2014}. We set the computational domain to be discretized with $\geq 10^4$ mesh nodes, and the relative numerical error tolerance is set to $< 10^{-8}$. The smallest frequency parameter considered in our computations is $\tilde{\omega} = 0.0001$. Although solutions for $\tilde{\omega} < 0.0001$ could theoretically be obtained by further refining the mesh, numerical convergence becomes increasingly challenging as the stiffness of the system increases.

\section{Numerical results}\label{sect4}


\begin{figure}
\centering
\includegraphics[width=0.48\linewidth]{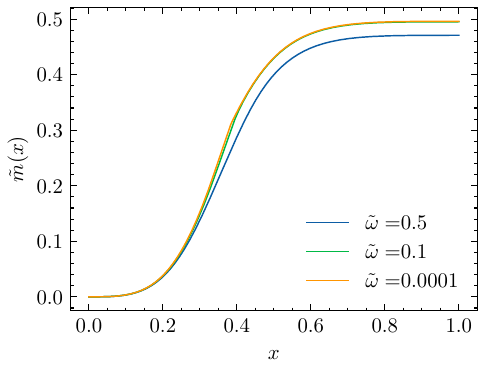}
\includegraphics[width=0.48\linewidth]{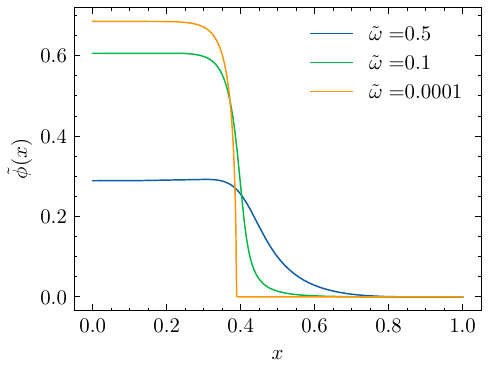}
\includegraphics[width=0.48\linewidth]{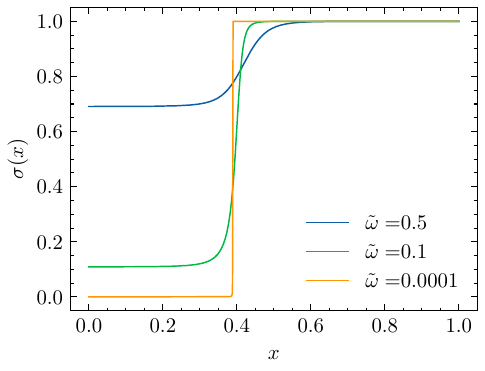}

\caption{Plots of $\tilde{m}(x)$, $\tilde{\phi}(x)$, and $\sigma(x)$ with parameters $\tilde{q}=0.56$, $s=0.2$, and $\tilde{\omega}=0.1$.
 \label{sols1}
}
\end{figure}
 
\subsection{{ Frozen States and quasi-horizon}} \label{FZS}

\begin{figure}
\centering
\subfigure[]{
\includegraphics[width=0.5\linewidth]{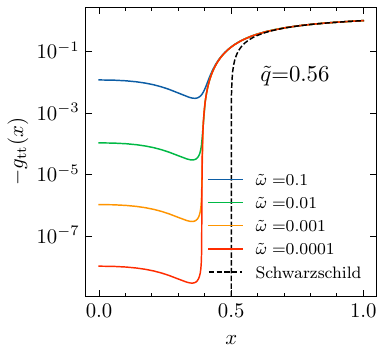}\label{fig61}
}
\subfigure[]{
\includegraphics[width=0.44\linewidth]{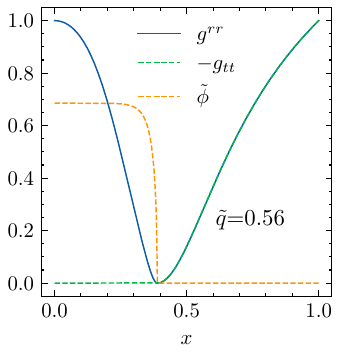}\label{fig62}
}

\caption{{(a)
Plots of metric components  $-g_{tt}$  with different values of  $\tilde\omega$.  (b) Plots of $-g_{tt}$, $g^{rr}$, and $\tilde{\phi}$ with   $\tilde\omega = 0.0001$. Other parameters are set to $\tilde q =0.56$ and $\tilde s =0.2$.} For reference, the black dashed line in the left panel corresponds to the case of  Schwarzschild black hole with mass $\tilde{M} =0.5$.
 \label{metric_n}
}
\end{figure}

{
For the Hayward-Boson star, when the magnetic monopole charge exceeds a certain critical value, denoted by $\tilde{q}_c$, the model permits solutions where $\tilde{\omega} \to 0$ ~\cite{YueWang2025}  (but $\tilde{\omega} \ne 0$\cite{FriedbergLeePang1987,Hod2018}, practically  down to $\tilde{\omega} = 0.0001$ ).

As shown in Fig.~\ref{sols1}, we demonstrate the variation of the numerical solution as the parameter $\tilde{\omega}$ gradually decreases, with other parameters $\tilde{q} = 0.56$ and $\tilde{s} = 0.2$ (the critical value $\tilde{q}_c = 0.49$). From the figure, it can be observed that when $\tilde{\omega} = 0.0001$, the value of the scalar field $\tilde{\phi}$ rapidly approaches zero near $x = 0.4$. This characteristic implies that we can define a soft boundary for the boson star, analogous to the hard boundary of a Schwarzschild black hole.

Moreover, as illustrated in Fig.~\ref{metric_n}a, as $\tilde{\omega}$ decreases, the minimum value of $-g_{tt}$ also decreases. When $\tilde{\omega}$ approaches $0.0001$, $-g_{tt}$ in the region centered at $x = 0$ drops significantly below $\mathcal{O}\  (10^{-7})$. Simultaneously, the minimum value of $g^{rr}$ also falls to $\mathcal{O}\ (10^{-7})$. To better confirm this soft boundary, we plot $-g_{tt}$, $g^{rr}$, and $\tilde{\phi}$ together, as shown in Fig.~\ref{metric_n}b. Where the mass of the Schwarzschild black hole used as a reference is $\tilde{M}=0.5$, which is the same as the ADM mass of the Hayward boson star when $\tilde{\omega} = 0.0001$, $\tilde{q} = 0.56$, and $\tilde{s} = 0.2$. Fig.~\ref{metric_n}b shows that the position of the minimum value of $g^{rr}$ (denoted as $x_c$) closely coincides with the positions where $-g_{tt}$ and $\tilde{\phi}$ rapidly approach to zero.
Therefore, we define the soft boundary of the boson star as the position of the minimum value of $g^{rr}$. Since $-g_{tt} \to 0$ at this soft boundary, we refer to it as the quasi-horizon, and refer to $x_c$ as the { quasi-horizon} radius}.

 {
As $x$ approaches $x_c$ from outside ($x \to x_c^+$), $-g_{tt} \to 0$. For a distant observer, this extreme gravitational redshift makes time appear to slow down asymptotically near $x_c$, such that matter approaching this surface from the outside cannot cross it within a finite coordinate time. This behavior aligns with the concept of a "frozen star" \cite{ZeldovichNovikov1971}.  
 Therefore, we refer to the boson star solutions that satisfy this condition as {\it frozen state} solutions. }

However, due to the limitations of numerical computation, we cannot obtain a solution where $-g_{tt}|_{x \to x_c} \to 0$ in practice. Therefore, we assume that when $\tilde{\omega} = 0.0001$ and $-g_{tt}|_{x \to x_c} \approx 10^{-7}$, the boson star is in a frozen state \cite{YueWang2025, HuangSunWang2025, ChenWang2024}. { And the  quasi-horizon} radius $\tilde{r}_{c}^\mathrm{H}$ in the physical coordinate system is then given by:
$$
\tilde{r}_{c}^\mathrm{H}  =  \frac{x_c}{1-x_c}.
$$

Furthermore, in the frozen state, the { quasi-horizon}  is not a truly  singularity. This can be determined by whether the Kretschmann scalar diverges. The expression for this scalar is  
\[  
K = R^{\rho \sigma \mu \nu} R_{\rho \sigma \mu \nu},  
\]  
where \(R^{\rho \sigma \mu \nu}$ and \(R_{\rho \sigma \mu \nu}$ are the contravariant and covariant forms of the Riemann curvature tensor, respectively. For the Schwarzschild solution, the expression for the Kretschmann scalar is:
\[  
K = \frac{48M^2}{r^6}.
\]
  
\begin{figure}
\centering
\includegraphics[width=0.7\linewidth]{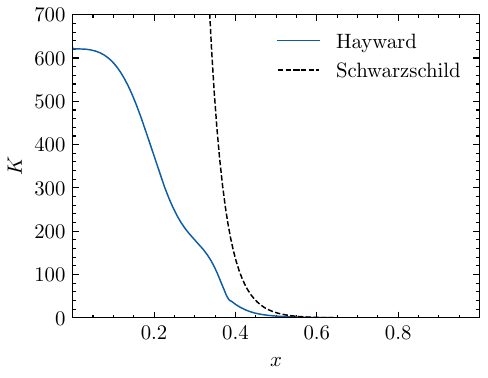}
\caption{Plots of Kretschmann scalar as a function of $x$ with $\tilde q =0.56$, $\tilde{\omega} =0.0001$, and $\tilde s =0.2$. For reference, the black dashed line  corresponds to the case of  Schwarzschild black hole with mass $\tilde{M} =0.5$..
\label{K}
}
\end{figure}

In Fig. \ref{K}, we present the variation of the Kretschmann scalar with $x$ {, where parameters are set to } $\tilde q = 0.56$, $\tilde{\omega} = 0.0001$, and $\tilde s = 0.2$. For reference,  the Kretschmann scalar for a Schwarzschild black hole with mass $\tilde{M} =0.5$ is also plotted. It can be found from the figure that the Kretschmann scalar does not exhibit any points of divergence for Hayward boson stars. Therefore,  there is no true singularity at the { quasi-horizon}, but only a coordinate singularity.

\subsection{Light Rings} \label{LRS}

The motion of particles in a gravitational field must satisfy the following equation:
\begin{equation}\label{geodesic}
g_{\mu\nu} \dot{x}^{\mu} \dot{x}^{\nu} = - \epsilon,
\end{equation}
where the dot denotes differentiation with respect to the affine parameter (for photons) or proper time (for massive particles) along the geodesic. Here, $x^{\mu} = (\tilde{t}, \tilde{r}, \theta, \varphi)$ represent the coordinate components, with $\tilde{t} = t/\mu$. The parameter $\epsilon = 0$ corresponds to photons, and $\epsilon = 1$ corresponds to massive particles. 

The particles' orbits are assumed to lie in the equatorial plane $\theta = \pi/2$, implying $\dot{\theta} = 0$.  Equation~\eqref{geodesic} then reduces to 
\begin{equation}\label{geodesic1}
\frac{1}{2} \dot{\tilde r}^2 + V(\tilde r) = 0, 
\end{equation}
with 
\begin{equation}\label{potential1}
V(\tilde r) = \frac{1}{2} \frac{1}{g_{tt}g_{rr}} \left( E^2+  g_{tt}\left(\epsilon+\frac{L^2}{\tilde r^2}\right) \right),
\end{equation}
Here, $L$ and $E$ are two constants of motion, defined as $L = \tilde{r}^2 \dot{\varphi}$ and $E = - g_{tt} \dot{\tilde{t}}$. While for photons, $L$ and $E$ correspond to their angular momentum and energy (defined up to an arbitrary affine scaling), for massive particles, these quantities represent the {specific angular momentum} and {specific energy} (i.e., angular momentum and energy per unit mass), respectively.

 Given the angular momentum $L$ and energy $E$, the conditions for a photon to undergo stable circular motion are:
\begin{equation}
V(\tilde r) = 0,\quad V'(\tilde r) = 0,\quad \text{and} \quad V''(\tilde r) > 0.\label{cds1}
\end{equation}
Here, the conditions $V'(\tilde r) = 0$ and $V''(\tilde r) > 0$ are respectively equivalent to
\[
\left( -\frac{g_{tt}}{\tilde r^2} \right)' = 0, \quad \left( -\frac{g_{tt}}{\tilde r^2} \right)'' > 0.
\]
Therefore, we can define the following effective potential:
\begin{equation}
\label{potential3}
V_{\mathrm{eff}}^{\mathrm{pho}}(\tilde r) = -\frac{g_{tt}}{\tilde r^2},
\end{equation}
to assist in determining the stable circular orbits of photons, and modify the conditions in equation \ref{cds1} as follows:
\[
V(\tilde r) = 0,\quad V_{\mathrm{eff}}^{\mathrm{pho}}{}'(\tilde r) = 0,\quad \text{and} \quad V_{\mathrm{eff}}^{\mathrm{pho}}{}''(\tilde r)  {\ge}  0.
\]   
The stability of the orbit along the $\theta$-direction of  boson stars has been rigorously proved by Cunha, Berti and Herdeiro \cite{CunhaBertiHerdeiro2017}.

\begin{figure}
\centering
\includegraphics[width=0.7\linewidth]{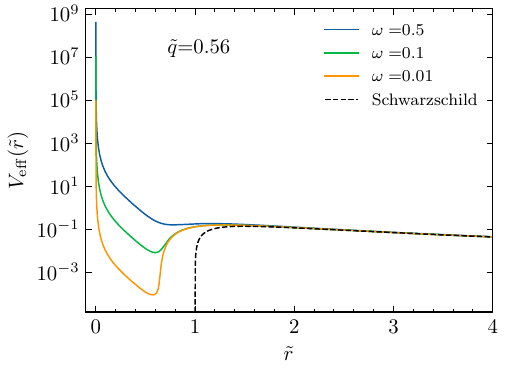} 
\label{V-LR-1b}

\caption{Effective potentials for photon circular orbits with  $\tilde q =0.56$ and $\tilde s =0.2$.  For reference, the black dashed line  corresponds to the case of  Schwarzschild black hole with mass $\tilde{M} =0.5$.
} 
\label{V-LR-1}
\end{figure}

Fig.~\ref{V-LR-1} shows the effective potential curves for different values of $\tilde{\omega}$ with  $\tilde{q} = 0.56$ and $\tilde{s} = 0.2$. Each effective potential curve exhibits two extreme points, {indicating the presence of two light rings (LRs)} \cite{MacedoPaniCardoso2013}. The location of these extrema shows that the inner light ring is stable, whereas the outer one is unstable. { This is significantly different from the case of a Schwarzschild black hole, where only one unstable light ring exists.}

\begin{figure}[htb]

\centering
\includegraphics[width=0.7\linewidth]{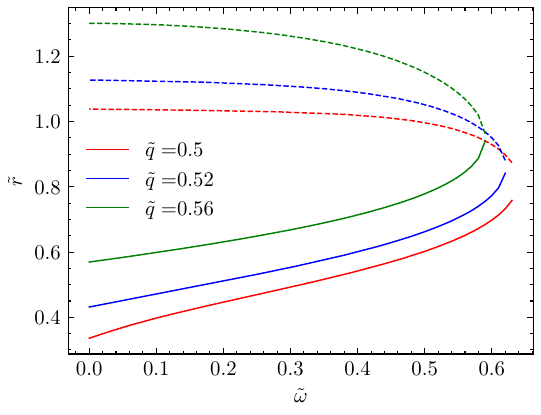}  
\label{LRs-a}

\caption{ {{ $\tilde{r}_\mathrm{inner}^\mathrm{LR}$ (solid lines) and $\tilde{r}_\mathrm{outer}^\mathrm{LR}$ (dashed lines)  versus $\tilde\omega$ with  $\tilde s = 0.2$.} }}
\label{LRs}
\end{figure}

The $\tilde{\omega}$-dependence of these light ring (LR) radii are shown in Fig.~\ref{LRs}, with different values of $\tilde{q}>\tilde{q}_c$. The inner and outer radii are denoted by $\tilde{r}_\mathrm{inner}^\mathrm{LR}$ and $\tilde{r}_\mathrm{outer}^\mathrm{LR}$, respectively. The figure reveals that { as $\tilde{\omega}$ increases, the values of $\tilde{r}_\mathrm{inner}^\mathrm{LR}$ and $\tilde{r}_\mathrm{outer}^\mathrm{LR}$ approach each other.}

\begin{figure}[htb]

\centering
\includegraphics[width=0.7\linewidth]{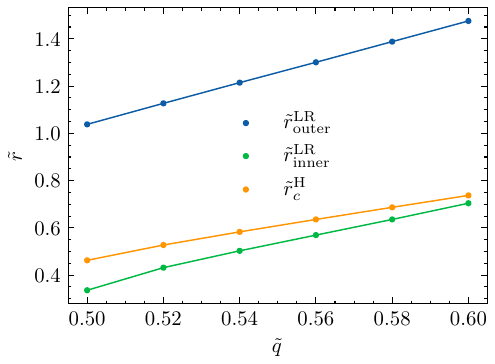}  

\caption{ {{  $\tilde r_\mathrm{c}^\mathrm{H}$, $\tilde r_\mathrm{inner}^{\mathrm{LR}}$ and $\tilde r_\mathrm{outer}^{\mathrm{LR}}$  versus $\tilde{q}$ for $\tilde \omega = 0.0001$  and $\tilde s = 0.2$. } }}
\label{LR-4}
\end{figure}

Furthermore, Fig.~\ref{LR-4} presents the relationship between the LR radii, the { quasi-horizon} radius and  $\tilde{q}$. The results show that as $\tilde{q}$ increases, the inner and outer LR radii and the { quasi-horizon} radius all increase monotonically. Notably, the inner light ring always lies within the { quasi-horizon} ($\tilde{r}_\mathrm{inner}^\mathrm{LR} < {\tilde r_\mathrm{c}^\mathrm{H}}$), while the outer light ring remains outside it ($\tilde{r}_\mathrm{outer}^\mathrm{LR} > { \tilde r_\mathrm{c}^\mathrm{H}}$).

\subsection{Accretion Disk and Shadows}

Here we assume that (1) the accretion disk is geometrically thin and optically thick \cite{NovikovThorne1973,PageThorne1974}, and (2) the observer's reference frame is at rest relative to the  star, satisfying the {distant} zero angular momentum observer (ZAMO) condition.

{{Similar to the photon case, the effective potential for massive particles can be given by}}
\begin{equation}
V_\mathrm{eff}^\mathrm{par}(\tilde r) = - g_{tt}\left(1+\frac{L^2}{\tilde r^2}\right).
\end{equation}
The conditions for stable circular motion require
\begin{equation}
V(\tilde r)=0, \quad V_\mathrm{eff}^\mathrm{par}{}'(\tilde r) = 0, \quad \text{and} \quad V_\mathrm{eff}^\mathrm{par}{}''(\tilde r) { \ge } 0.
\end{equation}
The inner radius of the accretion disk, denoted by $\tilde{r}_\mathrm{In}$, is defined as the radial coordinate where the second derivative of the effective potential $V_{\mathrm{eff}}^{\mathrm{par}}{''}(\tilde{r})$ first becomes zero as $\tilde{r}$ decreases. 

{ For the Schwarzschild black hole, the orbit corresponding to $V_{\mathrm{eff}}^{\mathrm{par}}{''}(\tilde{r})=0$ represents the innermost stable circular orbit [58]. 
However, the condition $V_{\mathrm{eff}}^{\mathrm{par}}{''}(\tilde{r})=0$ is not a sufficient condition for determining the innermost stable circular orbit (ISCO), and its use in determining the ISCO is highly model-dependent. For compact objects without event horizons (such as boson stars), the timelike circular orbits exhibit a richer variety of configurations [59], and this condition is generally not applicable. 
Here, we choose $\tilde{r}_\mathrm{In}$ as the inner edge of the accretion disk because the range of stable circular orbits extends continuously from $\tilde{r}_\mathrm{In}$ to radial infinity, for the convenience of comparison with the accretion disk of a Schwarzschild black hole.}

{{ The angular
velocity $\Omega$, the specific angular momentum  $L$ and the specific energy  $E$ for a massive particle in circular motion can be  determined by the following formulas:
\begin{equation}
\Omega =  \frac{\dd \varphi}{\dd {\tilde{t}}}  
=\pm \sqrt {\frac{- \partial_{\tilde{r}} g_{tt} }{\partial_{\tilde{r}} g_{\varphi\varphi}}},
\end{equation}}}
\begin{equation}
L =  \frac{g_{\varphi \varphi} \Omega}{\sqrt{- g_{tt} - g_{\varphi\varphi} \Omega^2}},
\end{equation}
\begin{equation}
E = - \frac{g_{tt} }{\sqrt{-g_{tt} - g_{\varphi\varphi} \Omega^2}}.
\end{equation}

%

Let $E_\mathrm{em}$ denote the energy of photons emitted from the accretion disk, and $E_\mathrm{obs}$ denote the energy received by the observer.
 Incorporating both gravitational redshift and Doppler effects, the redshift factor can be expressed as \cite{Luminet1979}

\begin{equation}\label{redshift}
1 + z = \frac{E_\mathrm{em}}{E_\mathrm{obs}} 
= \left.\frac{1 + \Omega b \sin{\theta_0}\cos{\alpha}}{\sqrt{-g_{tt} - g_{\phi\phi} \Omega^2}}\right|_{\tilde{r}=\tilde{r}_d},
\end{equation}  
where $\theta_0$, $\alpha$, and $\tilde{r}_d$ denote the inclination of the accretion disk, the angular coordinates of the pixels on the screen, and the radial coordinate of the particles on the disk, respectively.

Using the fixed observation frequencies of the Event Horizon Telescope (EHT) (1.3 mm \cite{AkiyamaAlberdiAlef2019e} or 0.87 mm \cite{RaymondDoelemanAsada2024}), we can invert  the emission spectrum distribution of the accretion disk through Equation \eqref{redshift}. Defining $\lambda_0$ as the observed wavelength and $\lambda$ as the emission wavelength from the disk, they satisfy:
\begin{equation}\label{lamd} 
\lambda = \frac{\lambda_0}{1 + z}.
\end{equation}

\begin{figure}[htb]
\centering
\includegraphics[width=\linewidth]{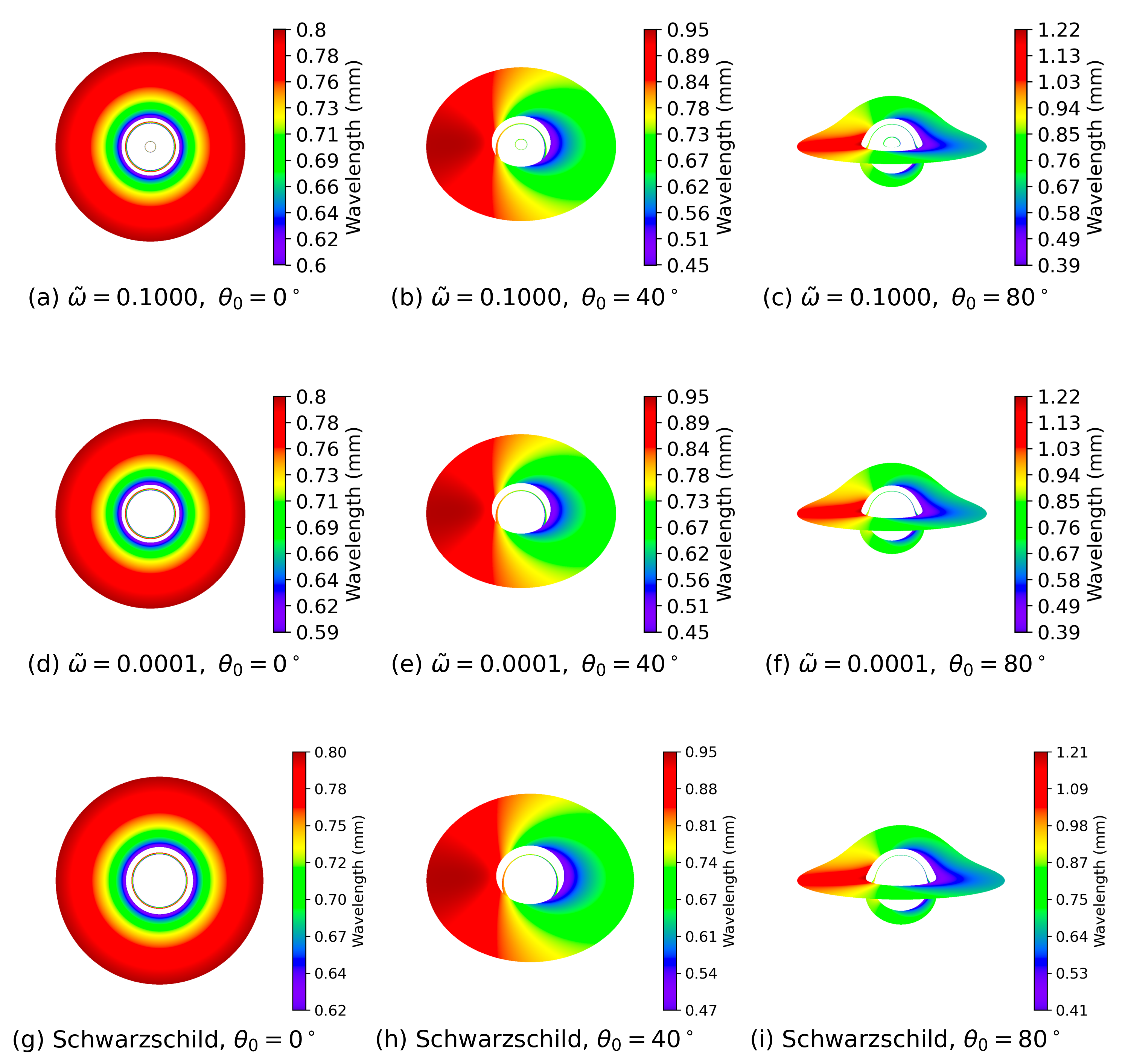}
\caption{
Spectral distribution and shadow   of the accretion disk with $\tilde{q}=0.56$. First row  $\tilde{\omega}=0.1$ and second row $\tilde{\omega}=0.0001$.   Columns from left to right correspond to inclination angles of $0^\circ$, $40^\circ$, and $80^\circ$ respectively, with disk radii ranging from 2.75 to 10 and $\lambda_0 = 0.87$~mm. {{The last row corresponds to the case of an { $\tilde{M}=0.5$} Schwarzschild black hole, where the accretion disk radius ranges from 3 to 10.}} To clearly display the photon rings inside the shadow region, we adjusted the background color to \emph{white}.
}
\label{disk_lmd0}
\end{figure}

Using the ray-tracing method and mapping the results to the 420-660 nm visible light range, we obtain the pseudo-color image of the accretion disk's spectral radiation distribution, as shown in Fig.~\ref{disk_lmd0} for $\lambda_0 = 0.87$~mm and $\tilde{q} = 0.56$. The inner and outer disk radii are set to $2.75$ and $10$, respectively. The distance from the boson star to the observer is $\tilde{r}_{\mathrm{obs}} = 100$, with a field of view (FOV) of $15^\circ$. {{The last row serves as a reference, corresponding to the case of a Schwarzschild black hole with { $\tilde{M}=0.5$} (this mass value is equivalent to the ADM mass of the boson star with $\tilde{\omega}=0.0001$ in the figure), where the accretion disk radius ranges from 3 to 10. }}  The background color is set to white to clearly reveal the secondary images (the \emph{photon rings}) within the central {shadow} region.

The first row in Fig.~\ref{disk_lmd0} shows the images at different inclination angles for $\tilde{\omega}=0.1$. At this value, the boson star is not in a frozen state. Compared to the accretion disk  image of a Schwarzschild black hole, an extra photon ring appears within the shadow region.  This phenomenon of multiple photon rings appearing in the shadow region has also been observed recently in other studies \cite{RosaMacedoRubiera-Garcia2023,GyulchevNedkovaVetsov2021,Tsupko2022,OlmoRubiera-GarciaSaez-ChillonGomez2022,Rosa2023,Gao2025}. Based on this characteristic, the observed object can be ruled out as a Schwarzschild black hole.

\begin{figure}[htb]
\centering
\includegraphics[width=1\linewidth]{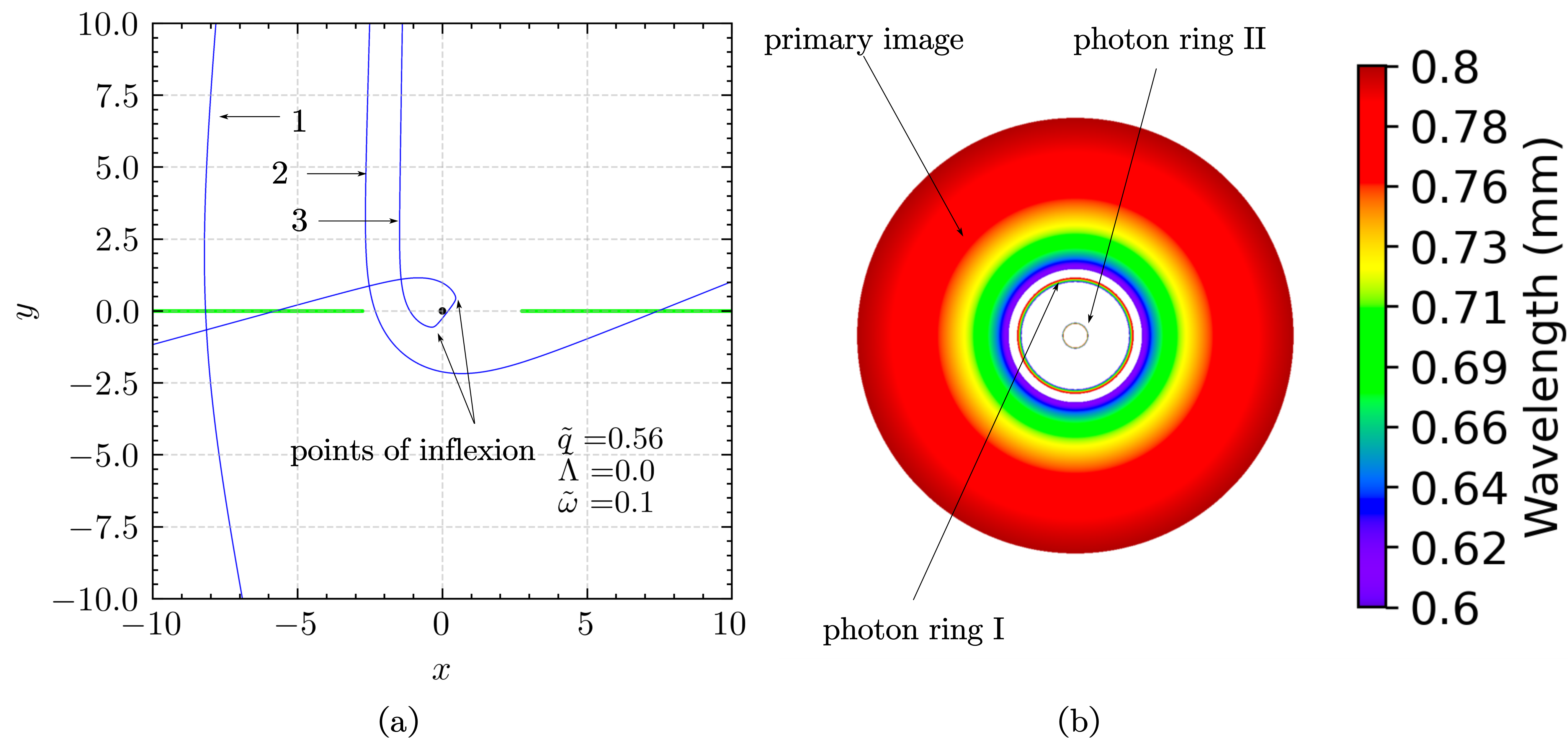}  
\caption{
(a) Photon trajectory cross-section ($\tilde{q}=0.56$,  $\tilde{w}=0.1$), where the green horizontal line indicates the accretion disk position, black dot marks the  star center. The incident angles of light rays 1, 2, and 3 are $5.0^\circ$, $1.6^\circ$ and $0.88^\circ$, respectively.  Light rays 1, 2, and  3 are the representative ones associated with  primary image, photon ring I, and photon ring II of (b),  respectively.
}
\label{tracing}
\end{figure}

To elucidate the image configuration of the Fig.~\ref{disk_lmd0}a, Fig.~\ref{tracing} illustrates the correspondence between photon trajectories and spectral images. 
In  Fig.~\ref{tracing}a, the ray 1 corresponds to the primary image, the ray 2 correspond to the photon ring I, and the ray 3 correspond to the  photon ring II.  It should be noted that the trajectory of  ray 3 exhibits double inflection points.

The second row of Fig.~\ref{disk_lmd0} presents results for $\tilde{\omega}=0.0001$. At this value, the boson star is in a frozen state. The presence of a { quasi-horizon} prevents light rays from crossing it, which results in the absence of a secondary photon ring (photon ring II). {{By comparing with the accretion disk structure of the Schwarzschild black hole in the third row of the figure, significant similarities can be observed.}}

%
%
%
%

\section{Conclusion}\label{Conc}

In this paper, we investigated some fundamental properties of the { Hayward boson stars with a magnetic monopole}, focusing on their frozen state solutions, light ring structures, shadow features, and observable spectral distributions from accretion disks. By numerically solving the coupled Einstein-nonlinear electrodynamics-scalar field equations, our main findings are as follows.

{
First, we confirm that the quasi-horizon of a frozen Hayward boson star is not truly singular but only a coordinate singularity.}

Furthermore, we examined the light ring structures. For { parameters  $\tilde{q}>\tilde{q}_c$. We find that} there can only be two light rings for different values of $\tilde\omega$. In the frozen state, there exists an inner stable light ring located within the { quasi-horizon} ($\tilde{r}_{\mathrm{inner}}^{\mathrm{LR}} < {\tilde r_\mathrm{c}^\mathrm{H}}$) and an outer unstable light ring outside the { quasi-horizon} ($\tilde{r}_{\mathrm{outer}}^{\mathrm{LR}} > {\tilde r_\mathrm{c}^\mathrm{H}}$). {As $\tilde{q}$ increases, the inner and outer LR radii and the { quasi-horizon} radius all increase monotonically.}

Finally, ray-tracing simulations  show that compared to a Schwarzschild black hole, the shadow region of a non-frozen Hayward   boson stars can contain an additional photon ring. We hope such as additional photon rings, may serve as an observational signature to distinguish such boson stars from black holes in future astronomical observations.  On the other hand, the absence of this additional ring in frozen Hayward boson stars enhances their resemblance to black holes, thereby strengthening their potential as black hole mimickers.

\section{Acknowledgment}
This work is supported by the National Natural Science Foundation of China (Grant No. 12275110  and No. 12247101)  and  the National Key Research and Development Program of China (Grant No. 2022YFC2204101 and 2020YFC2201503).



%
\biboptions{sort&compress}
{
\setstretch{0.9}                  
\let\oldthebibliography\thebibliography
\renewenvironment{thebibliography}[1]
{\begin{oldthebibliography}{#1}%
 \setlength{\itemsep}{0pt}
 \setlength{\parskip}{0pt}%
 \setlength{\parsep}{0pt}%
}
{\end{oldthebibliography}}
\bibliographystyle{elsarticle-num} 
\bibliography{library.bib}
}

\end{document}